\documentclass[conference]{IEEEtran}
\usepackage{graphicx}
\usepackage{amsmath}
\usepackage{array}
\newcommand{\PreserveBackslash}[1]{\let\temp=\\#1\let\\=\temp}
\newcolumntype{C}[1]{>{\PreserveBackslash\centering}p{#1}}
\newcolumntype{R}[1]{>{\PreserveBackslash\raggedleft}p{#1}}
\newcolumntype{L}[1]{>{\PreserveBackslash\raggedright}p{#1}}
\usepackage[usenames]{color}
\usepackage{colortbl,booktabs}
\usepackage{tabularx}
\usepackage{multicol}
\usepackage{booktabs}
\usepackage[Symbol]{upgreek}
\usepackage{subfigure}
\usepackage{bm}
\usepackage{cite}
\usepackage{balance}
\usepackage[boxed]{algorithm2e}

\usepackage{threeparttable}
\usepackage{amsmath}
\usepackage{dcolumn}
\usepackage{multirow}
\usepackage{booktabs}
\usepackage{amsfonts}
\newcolumntype{d}[1]{D{.}{.}{#1}}

\begin{document}

\bibliographystyle{IEEEtran} 
\title{Near-Optimal Hybrid Analog and Digital Precoding for Downlink mmWave Massive MIMO Systems}

\author{
\IEEEauthorblockN{Linglong Dai$^1$, Xinyu Gao$^1$, Jinguo Quan$^2$, and Shuangfeng Han$^3$, and Chih-Lin I$^3$}
\IEEEauthorblockA{$^1$Tsinghua National Laboratory for Information Science and Technology (TNList), \\Department of Electronic Engineering, Tsinghua University, Beijing, China\\
$^2$Division of Information Science $\&$ Technology, Shenzhen Graduate School, Tsinghua University, Shenzhen, China\\
$^3$Green Communication Research Center, China Mobile Research
Institute, Beijing 100053, China}}

\maketitle
\begin{abstract}
Millimeter wave (mmWave) massive MIMO can achieve orders of magnitude increase in spectral and energy efficiency, and it usually exploits the hybrid analog and digital precoding to overcome the serious signal attenuation induced by mmWave frequencies. However, most of hybrid precoding schemes focus on the full-array structure, which involves a high complexity. In this paper, we propose a near-optimal iterative hybrid precoding scheme based on the more realistic sub-array structure with low complexity. We first decompose the complicated capacity optimization problem into a series of ones easier to be handled by considering each antenna array one by one. Then we optimize the achievable capacity of  each antenna array from the first one to the last one by utilizing the idea of successive interference cancelation (SIC), which is realized in an iterative procedure that is easy to be parallelized. It is shown that the proposed hybrid precoding scheme can achieve better performance than other recently proposed hybrid precoding schemes, while it also enjoys an acceptable computational complexity.
\end{abstract}

\section{Introduction}\label{S1}

\IEEEPARstart The mergence of  millimeter-wave (mmWave) and massive multiple-input multiple-output (MIMO) is regarded as a promising technique for future 5G wireless communication systems~\cite{roh2014millimeter}, since it can provide orders of magnitude increase both in the available bandwidth and the spectrum efficiency~\cite{marzetta10,ngo11}. On one hand, the short wavelength associated with high frequencies of mmWave enables a large antenna array to be arranged in a small size. On the other hand, the large antenna array in massive MIMO can provide sufficient antenna gains to compensate for the serious signal attenuation induced by mmWave frequencies (e.g., rainfall effect and oxygen absorption~\cite{roh2014millimeter}) by exploiting the precoding technique to concentrate the signal in a specific direction.

However, the precoding schemes for massive MIMO are mainly performed in the baseband domain, where a fully digital precoder is utilized to eliminate the interferences by controlling both the amplitude and phase of transmitted signals. The fully digital precoding scheme, such as dirty paper precoding (DPC)~\cite{costa1983writing}, can achieve the optimal performance, but requires an expensive radio frequency (RF) chain (including digital-to-analog converter, up converter, and etc.) for every antenna. As a result, the total hardware complexity and energy consumption of digital precoding becomes a serious problem due to the increasing number of antennas, specially for mmWave communications where the number of antennas at the base station (BS) can be huge (e.g, 256 antennas can be used~\cite{roh2014millimeter}). To solve this problem, the hybrid precoding technique combining the digital precoding and the analog precoding has been proposed to significantly reduce the number of RF chains, is proposed~\cite{pi2011introduction}. As it can reduce both the hardware complexity and the energy consumption without obvious performance loss, it is considered as an essential technique for realistic mmWave massive MIMO systems~\cite{xiao2013suboptimal}. However, how to jointly design a near-optimal low-complexity digital and analog precoding is a challenging problem to be solved.

Recently, a hybrid precoding scheme, where the analog precoder is selected from the antenna array response vectors while the digital precoder is chosen from a predefined codebook~\cite{kim2013tens}, is proposed to achieve the satisfying performance, but it suffers from a high searching complexity. More recently, another scheme named spatially sparse precoding is proposed in~\cite{el2013spatially}. By formulating the capacity optimization problem as a sparse approximation problem, this scheme can approach the close-optimal capacity, but it involves very high computational complexity. Moreover, both of the schemes in~\cite{kim2013tens} and~\cite{el2013spatially} are designed for a full-array structure where each RF chain is connected to all BS transmit antennas, which makes the analog precoder too complicated in hardware. For a more realistic hybrid precoding structure where each RF chain is connected to only a small set of transmit antennas, i.e., the sub-array structure,~\cite{kim2013low} proposes a simple precoding scheme for a single data stream transmission, where the analog precoder and the digital precoder are selected from a predefined candidate set. By searching the optimal pair of analog and digital precoders, this scheme can achieve satisfying performance. For the multiple data streams transmission scenario, \cite{el2013multimode} proposes an iterative scheme where only the analog precoder is employed. It is simple and easy to be implemented in hardware, but suffers a non-negligible performance loss especially at low signal-to-noise (SNR) regions.

In this paper, we consider the mmWave massive MIMO systems with the sub-array structure for hybrid precoding and propose an iterative hybrid precoding scheme to achieve the near-optimal performance with low complexity. In particular, we decompose the complicated capacity optimization problem into a series of ones easier to be handled by considering each antenna array one by one, which means that the precoding for each antenna array connected to a specific RF chain will be considered separately instead of jointly. For the first antenna array, we use the digital precoder to control the amplitude, and the analog precoder to adjust the phase, and then the capacity optimization problem for the first antenna array will be solved with low complexity. After optimizing the achievable capacity of the first antenna array, we can utilize the idea of successive interference cancelation (SIC) to eliminate the contribution of the first antenna array from the total capacity expression, and then optimize the achievable capacity of the second antenna array. We repeat such procedure until the last antenna array is considered. Analysis and simulation results verify that the proposed hybrid precoding scheme can approach the ideal capacity with low complexity.

The rest of the paper is organized as follows. Section~\ref{S2} briefly introduces the mmWave massive MIMO system model. Section~\ref{S3} specifies the proposed near-optimal low-complexity precoding scheme, together with the complexity analysis. The simulation results of capacity performance are shown in Section~\ref{S4}. Finally, conclusions are drawn in Section~\ref{S5}.

{\it Notation}: Lower-case and upper-case boldface letters denote vectors and matrices, respectively;  ${( \cdot )^T}$, ${( \cdot )^H}$, ${( \cdot )^{ - 1}}$, and ${\det ( \cdot )}$  denote the transpose, conjugate transpose, inversion, and determinant of a matrix, respectively; ${{\left\|  \cdot  \right\|_2}}$  denotes the 2-norm of a vector; ${\left|  \cdot  \right|}$  denote the absolute operator; ${{\mathop{\rm Re}\nolimits} \{  \cdot \} }$ and ${{\mathop{\rm Im}\nolimits} \{  \cdot \} }$ denote the real part and imaginary part of a complex number, respectively; ${\mathbb{E}( \cdot )}$  denotes the expectation; Finally, ${{\bf{I}}_N}$ is the  $ N \times N $  identity matrix.

\section{System Model}\label{S2}
Fig. 1 compares the architecture of the fully digital precoding in conventional massive MIMO  and the hybrid analog/digital precoding in mmWave massive MIMO. In this paper, we consider the mmWave massive MIMO with the sub-array structure for hybrid precoding as shown in Fig. 1 (b), where the BS is equipped with ${NM}$  antennas but only ${N}$ independent RF chains to simultaneously transmit ${N}$ data streams for the ${K}$ receive antennas of user. The ${N}$ data streams in the baseband are firstly precoded by an ${N \times N}$  digital precoder ${{\bf{D}}}$, and then pass through ${N}$  RF chains. After that, each of the ${N}$ data streams is precoded again by an ${M \times 1}$ analog precoder ${{{\bf{a}}_i}}$ (${i = 1, \cdot  \cdot  \cdot N}$) before transmission, where the analog precoder is usually realized by a phase shifter~\cite{gholam2011beamforming}, i.e, all elements of ${{{\bf{a}}_i}}$  have the same amplitude of one but different phases. After analog precoding, each data stream is finally transmitted by ${M}$  antennas connected to the corresponding RF chain. Thus, the received ${K \times 1}$ signal vector ${{\bf{y}} = {[{y_1}, \cdot  \cdot  \cdot ,{y_K}]^T}}$  at the user side can be presented as
\begin{equation}\label{eq1}
{\bf{y}} = \rho {\bf{HADs}} + {\bf{n}} = \rho {\bf{HPs}} + {\bf{n}},
\end{equation}
where ${\rho }$ is the average received power, ${{\bf{s}} = {[{s_1}, \cdot  \cdot  \cdot ,{s_N}]^T}}$  presents the transmitted signal vector in the baseband, and usually ${\mathbb{E}({\bf{s}}{{\bf{s}}^H}) = \frac{1}{N}{{\bf{I}}_N}}$ is assumed for the normalized signal power~\cite{el2013spatially}. ${{\bf{P}} = {\bf{AD}}}$ presents the hybrid precoding matrix of size ${NM \times N}$. To reduce the complexity in hardware, the digital precoding matrix ${{\bf{D}}}$  can be designed as a diagonal matrix~\cite{han2014reference}, and the  ${i}$th diagonal elements ${{d_{i,i}}}$ can be a real constant. That means the digital precoder can be simply realized by an amplifier. ${{\bf{A}}}$  presents the ${NM \times N}$ analog precoding matrix as
\begin{equation}\label{eq2}
{\bf{A}} = {\left[ {\begin{array}{*{20}{c}}
{{{\bf{a}}_1}}&{\bf{0}}& \ldots &{\bf{0}}\\
{\bf{0}}&{{{\bf{a}}_2}}&{}&{\bf{0}}\\
 \vdots &{}& \ddots & \vdots \\
{\bf{0}}&{\bf{0}}& \ldots &{{{\bf{a}}_N}}
\end{array}} \right]_{NM \times N}}.
\end{equation}
where each row of ${{\bf{A}}}$ has only one nonzero entry due to the sub-array structure of the hybrid precoding scheme as shown in Fig. 1 (b). ${{\bf{H}} \in \mathbb{C}{^{K \times NM}}}$  denotes the channel matrix based on the Saleh-Valenzuela model~\cite{alkhateeb2013hybrid} widely used for mmWave communications. Finally, ${{\bf{n}} = {[{n_1}, \cdot  \cdot  \cdot ,{n_N}]^T}}$  is the additive white Gaussian noise (AWGN) vector, whose entries follow the independent and identically distribution (i.i.d.) ${{\cal C}{\cal N}(0,{\sigma ^2})}$.

\begin{figure}[h]
\begin{center}
\hspace*{0mm}\includegraphics[width=1\linewidth]{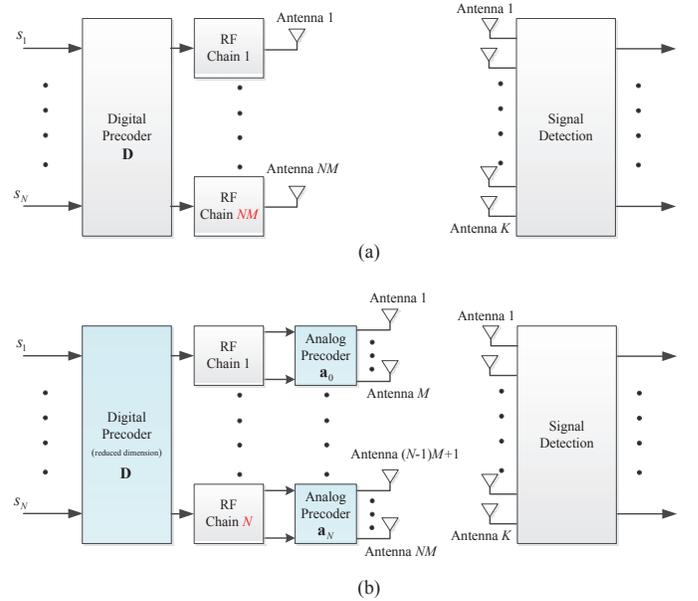}
\end{center}
\caption{Architecture comparison: (a) Fully digital precoding in conventional massive MIMO; (b) Hybrid analog/digital precoding in mmWave massive MIMO with sub-array structure.} \label{FIG1}
\end{figure}

It is known that mmWave channel will no longer obey the conventional Rayleigh fading due to the limited scatters~\cite{pi2011introduction,gaoel}. In this paper, we consider the geometric Saleh-Valenzuela channel model more appropriate for mmWave communications, where the channel matrix ${{\bf{H}}}$ can be presented as~\cite{alkhateeb2013hybrid}
\begin{equation}\label{eq3}
{\bf{H}} = \sqrt {\frac{{{N^2}M}}{L}} \sum\limits_{l = 1}^L {{\alpha _l}{\Lambda _r}} \left( {\phi _l^r} \right){\Lambda _t}\left( {\phi _l^t} \right){{\bf{f}}_r}\left( {\phi _l^r} \right){\bf{f}}_t^H\left( {\phi _l^t} \right),
\end{equation}
where ${L}$  is the number of channel paths corresponding to limited scatters, and we usually have ${L \le N}$ for  mmWave communication systems. ${{\alpha _l} \in \mathbb{C}}$ is the gain of the  ${l}$th path which follows the Rayleigh distribution. ${\phi _l^t}$ and ${\phi _l^r}$  are the azimuth angles of departure and arrival (AoDs/AoAs), respectively. ${{\Lambda _t}\left( {\phi _l^t} \right)}$  and ${{\Lambda _r}\left( {\phi _l^r} \right)}$  denote the transmit and receive antenna array gain at a specific AoD and AoA, respectively. For simplicity but without loss of generality, ${{\Lambda _t}\left( {\phi _l^t} \right)}$  and ${{\Lambda _r}\left( {\phi _l^r} \right)}$  can be set as one, which corresponds that both the receive and transmit antennas are assumed to be omni-directional~\cite{alkhateeb2013hybrid}. Finally, ${{{\bf{f}}_t}\left( {\phi _l^t} \right)}$  and  ${{{\bf{f}}_r}\left( {\phi _l^r} \right)}$ are the antenna array response vectors which heavily depends on the antenna array structure at the BS and the users, respectively. When both the transmit antenna array and the receive antenna array are the widely used uniform linear arrays (ULAs), we have
\begin{equation}\label{eq4}
{{\bf{f}}_r}\left( {\phi _l^r} \right)\! =\! \alpha{\left[ {1,{e^{j\frac{{2\pi }}{\lambda }d\sin \left( {\phi _l^r} \right)}}, \cdot  \cdot  \cdot ,{e^{j(N - 1)\frac{{2\pi }}{\lambda }d\sin \left( {\phi _l^r} \right)}}} \right]^T},
\end{equation}
\begin{equation}\label{eq5}
{{\bf{f}}_t}\left( {\phi _l^t} \right)\! =\! \beta{\left[ {1,{e^{j\frac{{2\pi }}{\lambda }d\sin \left( {\phi _l^t} \right)}}, \cdot  \cdot  \cdot ,{e^{j(NM - 1)\frac{{2\pi }}{\lambda }d\sin \left( {\phi _l^t} \right)}}} \right]^T},
\end{equation}
where ${\alpha  = \frac{1}{{\sqrt N }}}$ and ${\beta  = \frac{1}{{\sqrt NM }}}$ are the normalized factors, ${\lambda }$ denotes the wavelength of the signal, ${d}$ is the distance between two adjacent antenna elements. Note that in this paper, we only consider azimuth angles of the AoAs and AoDs without considering elevation angles, which means we will focus on the 2-D hybrid precoding scheme. The extensive design of 3D precoding scheme is also possible and will be left for further work.

\section{Near-Optimal Hybrid Analog/Digital Precoding}\label{S3}
In this section, we propose a near-optimal hybrid analog/digital precoding scheme by the joint design of analog and digital precoders for the mmWave massive MIMO system as illustrated in Fig. 1 (b). The complexity analysis is also provided to show its advantage over conventional schemes.
\subsection{The decomposition of the capacity optimization problem}\label{S3.1}
The final aim of precoding is to maximize the achievable channel capacity of MIMO systems, which can be expressed as~\cite{el2013spatially}
\begin{equation}\label{eq6}
R = {\log _2}\left( {\left| {{{\bf{I}}_N} + \frac{\rho }{{{\sigma ^2}}}{\bf{HP}}{{\bf{P}}^H}{{\bf{H}}^H}} \right|} \right).
\end{equation}
Based on the system model (1) as described in Section II, as the digital precoding matrix ${{\bf{D}}}$  is a diagonal matrix of real elements, there remains two constraints for the design of the hybrid precoding matrix ${{\bf{P}}}$: i) ${{\bf{P}}}$ should be a block diagonal matrix similar to the form of ${{\bf{A}}}$ as shown in (2); ii) All the non-zero elements of the  ${i}$th column of ${{\bf{P}}}$ has the same amplitude ${{d_{i,i}}}$, the ${i}$th diagonal element of the digital precoding matrix ${{\bf{D}}}$.

\begin{figure}[h]
\setlength{\abovecaptionskip}{0pt}
\setlength{\belowcaptionskip}{0pt}
\begin{center}
\hspace*{0mm}\includegraphics[width=0.9\linewidth]{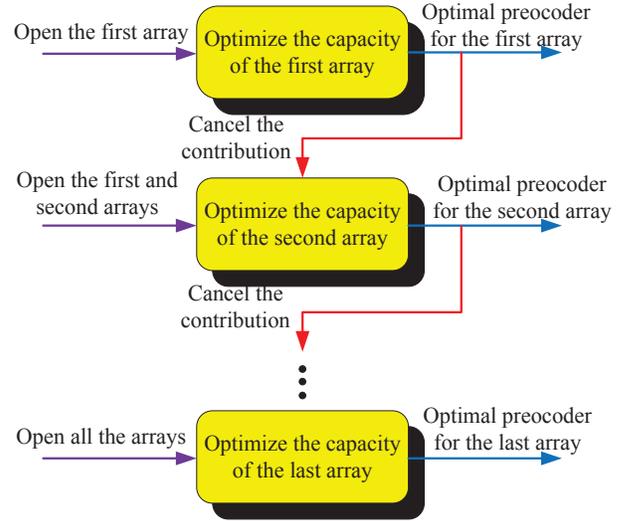}
\end{center}
\caption{Diagram of the proposed iterative hybrid analog/digital precoding.} \label{FIG2}
\end{figure}

Unfortunately, the non-convex constraints on ${{\bf{P}}}$  make the optimization problem~(\ref{eq6}) very difficult to be solved. To this end, as illustrated in Fig. 2, we propose to decompose the difficult optimization problem~(\ref{eq6}) into a series of sub-optimal problems much easier to be solved. In particular, by considering each antenna array connected to each RF chain one by one, we can  optimize the achievable capacity of the first antenna array by assuming that all the other antenna arrays are closed. After that, we can eliminate the contribution of the first antenna array from~(\ref{eq6}), and then optimize the achievable capacity of the second antenna array. Such similar procedure will be executed until the last antenna array is considered. The proposed iterative hybrid analog/digital precoding scheme will be described in detail in the next subsection.

\subsection{Near-optimal hybrid analog/digital precoding}\label{S3.2}
According to the analysis above, the achievable capacity of the first antenna array can be expressed by
\begin{equation}\label{eq7}
{C_1} = {\log _2}(1 + \frac{\rho }{{{\sigma ^2}}}{{\bf{p}}_1}^H{{\bf{H}}^H}{\bf{H}}{{\bf{p}}_1}),
\end{equation}
where ${{{\bf{p}}_1} = {d_{1,1}}{{\bf{b}}_1}}$ is the first column of the hybrid precoding matrix ${{\bf{P}}}$ (i.e., the precoder vector for the first antenna array), ${{{\bf{b}}_1}}$ is the first column of the analog precoding matrix ${{\bf{A}}}$  containing the ${M \times 1}$ non-zero vector ${{{\bf{a}}_1}}$ in (2) and the ${(N - 1)M \times 1}$ zero vector, i.e., ${{{\bf{b}}_1} = \left[ {{{\bf{a}}_1}\;{\bf{0}}} \right]_{NM \times 1}^T}$. Let ${{\bf{G}} = {{\bf{H}}^H}{\bf{H}}}$, since only the first ${M}$ elements of ${{{\bf{p}}_1}}$  are non-zeros, the term ${{{\bf{p}}_1}^H{\bf{G}}{{\bf{p}}_1}}$  which is required to be optimized in~(\ref{eq7}) can be rewritten as
\begin{equation}\label{eq8}
{{\bf{p}}_1}^H{\bf{G}}{{\bf{p}}_1} = {{\bf{p}}_1}^H{\bf{S}}{{\bf{p}}_1},
\end{equation}
where ${{\bf{S}}}$ is a sub-matrix of ${{\bf{G}}}$ which only keeps the rows and columns of ${{\bf{G}}}$  from 1 to ${M}$. Define the ordered singular value decomposition (SVD) of the Hermitian matrix  ${{\bf{S}}}$ as ${{\bf{S}} = {\bf{V\Sigma }}{{\bf{V}}^H}}$, where ${{\bf{\Sigma }}}$ is a ${M \times M}$  diagonal matrix containing the singular values of ${{\bf{S}}}$ in a decreasing order, and ${{\bf{V}}}$ is an ${M \times M}$ unitary matrix. It is known that the optimal solution to the objective~(\ref{eq7}) can be obtained by~\cite{golub2012matrix}
\begin{equation}\label{eq9}
{{\bf{p}}_{1,opt}} = {\left[ \begin{array}{l}
{{\bf{v}}_1}\\
{\bf{0}}
\end{array} \right]_{NM \times 1}},
\end{equation}
where the ${M \times 1}$  vector ${{{\bf{v}}_1}}$ is the first column of ${{\bf{V}}}$. However, based on the constraints as we mentioned in Section~\ref{S3}-A, we cannot directly choose ${{{\bf{p}}_1}}$  as ${{{\bf{p}}_{1,opt}}}$  since the elements of ${{{\bf{p}}_{1,opt}}}$ don't obey the constraint of constant amplitude. For this reason, we propose to explore a choice of  ${{{\bf{p}}_1}}$ that is sufficiently close to ${{{\bf{p}}_{1,opt}}}$  by minimizing the mean-square-error (MSE) between ${{{\bf{p}}_1}}$  and ${{{\bf{p}}_{1,opt}}}$  (or equivalently ${{d_{1,1}}{{\bf{a}}_1}}$  and ${{{\bf{v}}_1}}$) under the constraint of constant amplitude. Specifically, the MSE function can presented as
\begin{equation}\label{eq10}
\begin{array}{l}
\mathbb{E} \left\{ {\left\| {{{\bf{p}}_{1,opt}} - {{\bf{p}}_1}} \right\|_2^2} \right\}\\
\quad  = \mathbb{E} \left\{ {\left\| {{{\bf{v}}_1} - {d_{1,1}}{{\bf{a}}_1}} \right\|_2^2} \right\}\\
\quad  = \mathbb{E} \left\{ {{{\left( {{{\bf{v}}_1} - {d_{1,1}}{{\bf{a}}_1}} \right)}^H}\left( {{{\bf{v}}_1} - {d_{1,1}}{{\bf{a}}_1}} \right)} \right\}\\
\quad  = \mathbb{E} \left\{ {{\bf{v}}_1^H{{\bf{v}}_1} - {d_{1,1}}({\bf{a}}_1^H{{\bf{v}}_1} + {\bf{v}}_1^H{{\bf{a}}_1}) + d_{1,1}^2{\bf{a}}_1^H{{\bf{a}}_1}} \right\}.
\end{array}
\end{equation}
According to facts that the elements of ${M \times 1}$ non-zero vector ${{{\bf{a}}_1}}$ only have different phase and ${{d_{1,1}}}$ is a real scalar, a feasible and intuitive choice of ${{{\bf{a}}_1}}$ is
\begin{equation}\label{eq11}
{{\bf{a}}_1} = {e^{j \rm{angle}({{\bf{v}}_1})}},
\end{equation}
where ${{\rm{angle(}}{{\bf{v}}_1}{\rm{)}}}$ denotes the phase vector of ${{{\bf{v}}_1}}$, which means that the element of ${{{\bf{a}}_1}}$  will share the same phase as the corresponding element of ${{{\bf{v}}_1}}$  with amplitude of one. Substituting (\ref{eq11}) into (\ref{eq10}), we can observe that the MSE function (\ref{eq10})  will be simplified to a quadratic function that only depends on ${{d_{1,1}}}$  as
\begin{equation}\label{eq12}
\begin{array}{l}
\mathbb{E} \left\{ {\left\| {{{\bf{p}}_{1,opt}} - {{\bf{p}}_1}} \right\|_2^2} \right\}\\
\quad =\! \mathbb{E} \left\{ {{\bf{v}}_1^H{{\bf{v}}_1} - {d_{1,1}}({\bf{a}}_1^H{{\bf{v}}_1} + {\bf{v}}_1^H{{\bf{a}}_1}) + d_{1,1}^2{\bf{a}}_1^H{{\bf{a}}_1}} \right\}\\
\quad =\! 1\! +\! \mathbb{E} \left\{Md_{1,1}^2\! -\! {{d_{1,1}}\left( {{e^{j \rm{angle}({\bf{v}}_1^H)}}{{\bf{v}}_1}\! +\! {\bf{v}}_1^H{e^{j \rm{angle}({{\bf{v}}_1})}}} \right)} \right\}.
\end{array}
\end{equation}
where ${{\bf{a}}_1^H{{\bf{a}}_1} = \left\| {{e^{j \rm{angle}({{\bf{v}}_1})}}} \right\|_2^2 = M}$, ${{\bf{v}}_1^H{{\bf{v}}_1} = 1}$ since ${{{\bf{v}}_1}}$ is a column of unitary matrix ${{\bf{V}}}$. To minimize~(\ref{eq12}), it is easy to obtain the best choice of ${{d_{1,1}}}$ as
\begin{equation}\label{eq13}
{d_{11}} = \frac{{{e^{j \rm{angle}({\bf{v}}_1^H)}}{{\bf{v}}_1} + {\bf{v}}_1^H{e^{j \rm{angle}({{\bf{v}}_1})}}}}{{2M}} = \frac{{{\mathop{\rm Re}\nolimits} \left[ {{\bf{v}}_1^H{e^{j \rm{angle}({{\bf{v}}_1})}}} \right]}}{M}.
\end{equation}
Combining~(\ref{eq11}) and~(\ref{eq13}), we can obtain the precoder vector ${{{\bf{p}}_1}}$ for the first antenna array sufficiently close to ${{{\bf{p}}_{1,opt}}}$  as
\begin{equation}\label{eq14}
{{\bf{p}}_1} = {\left[ \begin{array}{l}
\frac{{{\mathop{\rm Re}\nolimits} \left[ {{\bf{v}}_1^H{e^{j \rm{angle}({{\bf{v}}_1})}}} \right]}}{M} \times {e^{j \rm{angle}({{\bf{v}}_1})}}\\
\quad \quad \quad \quad \quad \quad {\bf{0}}
\end{array} \right]_{NM \times 1}}.
\end{equation}

After acquiring the precoder vector for the first antenna array, we can utilize the idea of SIC to eliminate the contribution of the first antenna array from the total channel capacity~(\ref{eq6}), and then focus on the optimal design of the second precoder vector, and so on. Specifically, when the precoder vectors for the first ${m}$ antenna arrays have been obtained, by exploiting the Sherman-Morrison determinant identity~\cite{el2013multimode}, the total channel capacity ${R}$ after adding the contribution of the  (${m+1}$)th antenna array should be
\begin{equation}\label{eq15}
R\! =\! {\log _2}\left( {\left| {\bf{T}}_m \right|} \right)\! +\! {\log _2}\left\{ {1 + \frac{\rho }{{{\sigma ^2}}}{\bf{p}}_{m + 1}^H{{\bf{H}}^H}{{\bf{T}}_m^{ - 1}}{\bf{H}}{{\bf{p}}_{m + 1}}} \right\},
\end{equation}
\begin{equation}\label{eq16}
{\bf{T}}_m = {{\bf{I}}_N} + \frac{\rho }{{{\sigma ^2}}}{\bf{HP}}\left( {1:m} \right){\bf{P}}{\left( {1:m} \right)^H}{{\bf{H}}^H},
\end{equation}
where ${{\bf{P}}\left( {1:m} \right)}$ denotes the first ${m}$ columns of ${{\bf{P}}}$  and ${{{\bf{p}}_{m + 1}} = {d_{m + 1,m + 1}}{{\bf{b}}_{m + 1}}}$ is the (${m+1}$)th column of  ${{\bf{P}}}$  which corresponds to the precoder vector for the (${m+1}$)th antenna array. We can observe that the first term on the right side of (\ref{eq15}) is independent of ${{{\bf{p}}_{m + 1}}}$. Therefore, maximizing (\ref{eq15}) as a function of  ${{{\bf{p}}_{m + 1}}}$ equals to maximizing the second term of (\ref{eq15}) on the right side. Note that this term and (\ref{eq7}) share a similar form. Therefore, if we redefine the matrix ${{\bf{G}}}$  as
\begin{equation}\label{eq17}
{\bf{G}} = {{\bf{H}}^H}{\bf{T}}_m^{ - 1}{\bf{H}},
\end{equation}
we can utilize the similar method when we consider  ${{{\bf{p}}_1}}$ in (7) to obtain a near-optimal precoder vector ${{{\bf{p}}_{m + 1}}}$ for the (${m+1}$)th antenna array. After ${{{\bf{p}}_{m + 1}}}$ has been obtained, the precoder vector for the rest antenna arrays can be also acquired similarly.

\begin{algorithm}[h]
\caption{Iterative hybrid analog/digital preocoding}
\KwIn{(1) Number of RF chains ${N}$;
\\\hspace*{+9.3mm} (2) Number of antennas for one RF chain ${M}$;
\\\hspace*{+9.3mm} (3) Channel matrix ${{\bf{H}}}$}
 \textbf{Initialization}: ${{\bf{P}} = {{\bf{0}}_{NM \times N}}}$

 \textbf{for} ${1 \le m \le N}$
 \\\hspace*{+2.5mm} 1) Update matrix ${{\bf{G}}}$:
 \\\hspace*{+7mm} ${{\bf{G}} \leftarrow {{\bf{H}}^H}{\bf{T}}_m^{ - 1}{\bf{H}}}$;
 \\\hspace*{+2.5mm} 2) Acquire the effective sub-matrix of ${{\bf{G}}}$:
 \\\hspace*{+7mm} ${{\bf{S}} \leftarrow {\bf{G}}(M(m - 1):Mm,M(m - 1):Mm)}$;
 \\\hspace*{+2.5mm} 3) SVD:
 \\\hspace*{+7mm} ${{{\bf{v}}_1} \leftarrow }$ The first right-singular vector of  ${{\bf{S}}}$;
 \\\hspace*{+2.5mm} 4) The analog precoder:
 \\\hspace*{+7mm} ${{{\bf{a}}_m} \leftarrow {e^{j{\rm{angle}}({{\bf{v}}_1})}}}$, ${{{\bf{b}}_m} \leftarrow {\left[ \begin{array}{l}
{{\bf{0}}_{M(m - 1) \times 1}}\\
{{\bf{a}}_m}\\
{{\bf{0}}_{M(N - m) \times 1}}
\end{array} \right]}}$;
 \\\hspace*{+2.5mm} 5) The digital precoder:
 \\\hspace*{+7mm} ${{d_{m,m}} \leftarrow \frac{{{\mathop{\rm Re}\nolimits} \left[ {{\bf{v}}_1^H{{\bf{a}}_m}} \right]}}{M}}$;
 \\\hspace*{+2.5mm} 6) The total precoder:
 \\\hspace*{+7mm} ${{{\bf{p}}_m} \leftarrow {d_{m,m}}{{\bf{b}}_m}}$;

\textbf{end for}

\KwOut{(1) ${{\bf{A}} = \left[ {{{\bf{b}}_1},{{\bf{b}}_2}, \cdot  \cdot  \cdot ,{{\bf{b}}_N}} \right]}$;
\\\hspace*{+12mm} (2) ${{\bf{D}} = {\rm{diag}}({d_{1,1}},{d_{2,2}}, \cdot  \cdot  \cdot ,{d_{N,N}})}$;
\\\hspace*{+12mm} (3) ${{\bf{P}} = \left[ {{{\bf{p}}_1},{{\bf{p}}_2}, \cdot  \cdot  \cdot ,{{\bf{p}}_N}} \right]}$.}
\end{algorithm}

To sum up, the pseudo-code of the proposed iterative hybrid analog/digital preocoding scheme is described in \textbf{Algorithm 1}. Note that this scheme can be guaranteed to achieve a near-optimal performance, which is close to the upper-bound achieved by the optimal scheme when we choose ${{{\bf{p}}_m} = {{\bf{v}}_1}}$ without considering the constraint of constant amplitude, as will be verified later by the simulation results in Section~\ref{S4}.

\subsection{Complexity analysis}\label{S3.3}
In this subsection, we compare the computational complexity of the proposed iterative hybrid analog/digital precoding and the fully analog precoding in~\cite{el2013multimode}. Since both of them need to compute (\ref{eq17}) in each iteration, we compare their complexity after we have obtained the matrix ${{\bf{G}}}$ in terms of the required number of multiplications~\cite{gao2014low}.

It can be found from \textbf{Algorithm 1} that the complexity of the proposed scheme except computing (\ref{eq17}) comes from four parts. The first one is the SVD of the ${M \times M}$  matrix ${{\bf{S}}}$ to obtain ${{{\bf{v}}_1}}$. There are several methods to realize SVD such as the one based on QR decomposition, or the one based on Jacobi eigenvalue algorithm~\cite{golub2012matrix}, and the corresponding complexity is ${{\cal O}({M^3})}$~\cite{golub2012matrix}. The second one is to obtain the analog precoder ${{{\bf{a}}_m}}$. Since ${{{\bf{a}}_m}}$  only presents the phase of   ${{{\bf{v}}_1}}$ with amplitude of one, this part involves no extra computation. The third one comes from the digital precoder ${{d_{m,m}}}$. This part requires one multiplication of the ${1 \times M}$  vector ${{\bf{v}}_1^H}$  and the ${M \times 1}$  vector ${{{\bf{a}}_m}}$. Therefore, the complexity is ${{\cal O}({M^2})}$. The last one originates from computing the hybrid precoding matrix ${{{\bf{p}}_m}}$. Since there are only ${M}$  non-zero elements in ${{{\bf{p}}_m}}$, the complexity of this part is ${{\cal O}(M)}$.

To sum up, the overall complexity of the proposed iterative hybrid precoding scheme in each iteration is ${{\cal O}({M^3})}$.

In contrast, the computational complexity of the fully analog precoding scheme~\cite{el2013multimode} in each iteration is ${{\cal O}(N{M^2})}$, which means the computational complexity of the proposed scheme is comparable with that of the fully analog precoding scheme.  Additionally, the process of each iteration of the fully analog precoding scheme is hard to be parallelized~\cite{el2013multimode}, while our method can be easily implemented in a parallel manner, which indicates that lower hardware complexity can be achieved by the proposed iterative hybrid analog/digital precoding scheme.

\section{Simulation Results}\label{S4}
To evaluate the performance of the proposed iterative hybrid analog/digital precoding scheme, we provide the simulation results of the achievable channel capacity. Here we also provide the performance of the recently proposed fully analog precoding scheme~\cite{el2013multimode} based on the sub-array structure and the optimal precoder utilizing the precoding vector ${{{\bf{p}}_m} = {{\bf{v}}_1}}$ as the benchmark for comparison. The simulation parameters are described as follows. The carrier frequency is set as 28GHz~\cite{roh2014millimeter}, and two typical mmWave massive MIMO configurations with ${NM \times K = 128 \times 16}$ (${M = 8}$) and ${NM \times K = 128 \times 32}$ (${M = 4}$) are considered, respectively. Both the transmit and receive antenna arrays are ULAs with antenna spacing ${d = \lambda /2}$. We generate the channel matrix according to the channel model~\cite{alkhateeb2013hybrid} described in Section~\ref{S2}. The AoAs/AoDs are assumed to follow the uniform distribution within ${[0,2\pi ]}$. The number of scattering propagation paths is set as ${L = 10}$.

\begin{figure}[h]
\setlength{\abovecaptionskip}{0pt}
\setlength{\belowcaptionskip}{0pt}
\begin{center}
\vspace*{1mm}\includegraphics[width=0.9\linewidth]{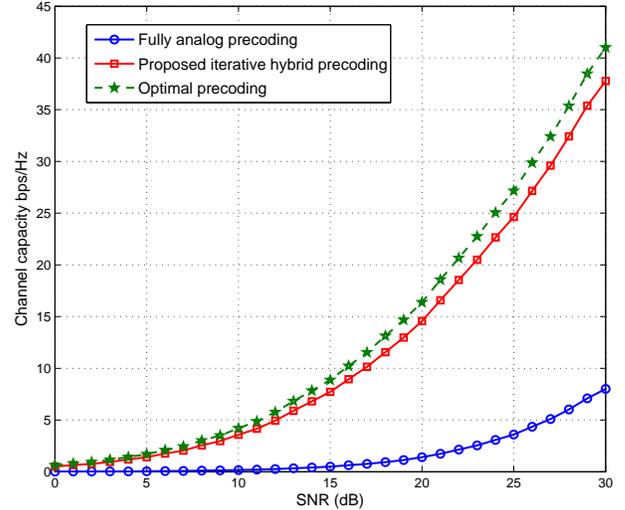}
\end{center}
\caption{Capacity comparison for an ${NM \times K = 128 \times 16}$ (${M = 8}$) mmWave massive MIMO system.} \label{FIG3}
\end{figure}

\begin{figure}[h]
\setlength{\abovecaptionskip}{0pt}
\setlength{\belowcaptionskip}{0pt}
\begin{center}
\hspace*{0mm}\includegraphics[width=0.9\linewidth]{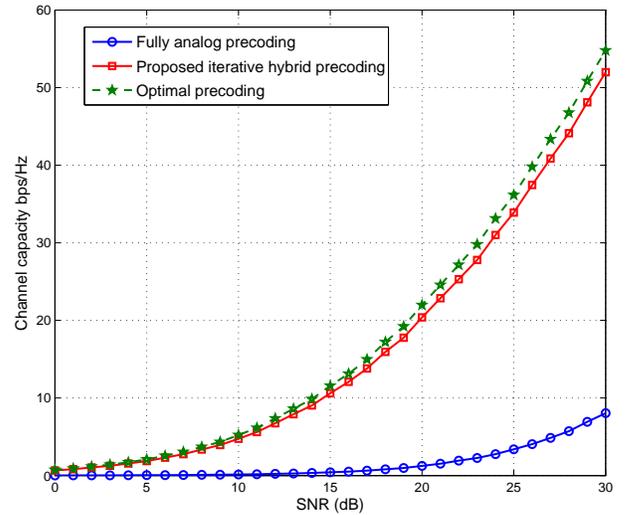}
\end{center}
\caption{Capacity comparison for an ${NM \times K = 128 \times 32}$ (${M = 4}$) mmWave massive MIMO system.} \label{FIG4}
\end{figure}

Fig. 3 shows the capacity comparison between the fully analog precoding scheme~\cite{el2013multimode} and the proposed iterative hybrid precoding scheme when ${NM \times K = 128 \times 16}$ (${M = 8}$). Note that~\cite{el2013multimode} also proposes a multimode strategy, where each antenna array can provide additional gain for the existing data stream instead of transmitting a new independent data stream. However, since the performance gain of this strategy is negligible at high SNR (${ > 5{\rm{dB}}}$)~\cite{el2013multimode}, here we only consider that each antenna array transmits an independent data stream. We can observe from Fig. 3 that the proposed iterative hybrid precoding scheme outperforms the fully analog one in the whole SNR range, and the performance gap becomes more obvious with the increasing SNR. For example, when SNR = 10 dB, the  capacity gap between the proposed scheme and the conventional one is about 3 bps/Hz, while when SNR = 30 dB, the capacity gap increases to 30 bps/Hz. What's more, Fig. 3 also shows that the proposed scheme is near-optimal, since the required SNR gap between the optimal precoding and the proposed one to achieve the same capacity is within 1 dB.

Fig. 4 shows the capacity comparison in the mmWave massive MIMO system with ${NM \times K = 128 \times 32}$ (${M = 4}$). Comparing Fig. 3 and Fig. 4, we can find that with the increasing number of RF chains (i.e., ${N}$ is increased from 16 to 32), the performance of the fully analog precoding becomes worse. For example, for the ${NM \times K = 128 \times 16}$  system, when SNR = 30 dB, the fully analog scheme can achieve ${20\% }$ of the optimal capacity, while for the  ${NM \times K = 128 \times 32}$ system, it can only achieve ${16\% }$ of the optimal capacity. In contrast, the proposed iterative precoding scheme can achieve  ${88\% }$ and ${96\% }$ of the optimal capacity for the ${NM \times K = 128 \times 16}$  and ${NM \times K = 128 \times 32}$ systems, respectively. This can be explained by the fact that the performance improvement obtained by the digital precoder becomes obvious with the increasing number of RF chains, and the fully analog precoder system will suffer from a non-negligible performance loss. More importantly, we can also conclude that when the number of antennas connected to each RF chain (i.e., ${M}$) decreases, the proposed scheme will be more close to the optimal precoding scheme.

\vspace*{+5mm}
\section{Conclusions}\label{S5}
In this paper, we propose a near-optimal hybrid analog/digital precoding scheme with low complexity for mmWave massive MIMO systems. By considering each antenna array separately and exploiting the idea of SIC, we decompose the complicated capacity optimization problem into a series sub-optimal problems much easier to be solved, and realize the proposed scheme in an iterative way which can be implemented in a parallel manner. It shows that the computational complexity of the proposed scheme is ${{\cal O}({M^3})}$, which is comparable with the conventional fully analog precoding scheme. Simulation results verify that the performance of the proposed scheme is close to the optimal capacity, especially when ${M}$ is relatively small (e.g., ${M=4}$). Our further work will focus on extending the proposed hybrid precoding scheme from 2D scenario to 3D scenario, where the elevation angles of the AoAs/AoDs will be also taken into account.

\balance

\vspace*{+5mm}
\section*{Acknowledgments}
This work was supported by National Key Basic Research Program of China (Grant No. 2013CB329203), National High Technology Research and Development Program of China (Grant No. 2014AA01A704), National Nature Science Foundation of China (Grant Nos. 61271266 and 61201185), Science and Technology Foundation for Beijing Outstanding Doctoral Dissertation Supervisor (Grant No. 20121000303), and Foundation of Shenzhen government.

\vspace*{-3mm}
\bibliography{IEEEabrv,Gao1Ref}

\end{document}